\documentclass[doublecol]{epl2}

\usepackage{calrsfs}

\title{Skyrmion soliton motion on periodic substrates by atomistic and particle based simulations}

\shorttitle{Skyrmion soliton motion on periodic substrates}

\author{J. C. B. Souza\inst{1} \and N. P. Vizarim\inst{1} \and C. J. O. Reichhardt\inst{2} \and C. Reichhardt\inst{2} \and P. A. Venegas\inst{3}}

\shortauthor{J. C. B. Souza \etal}

\institute{
    \inst{1} POSMAT - Programa de P\'os-Gradua\c{c}\~ao em Ci\^encia e Tecnologia de Materiais, S\~ao Paulo State University (UNESP), School of Sciences, Bauru 17033-360, SP, Brazil\\
    \inst{2} Theoretical Division and Center for Nonlinear Studies, Los Alamos National Laboratory, Los Alamos, New Mexico 87545, USA\\
    \inst{3} Department of Physics, S\~ao Paulo State University (UNESP), School of Sciences, Bauru 17033-360, SP, Brazil
}

\date{\today}

\abstract{
We compare the dynamical behavior of magnetic skyrmions
interacting with square and triangular defect arrays just
above commensuration using both an atomistic model and a
particle-based model. Under an applied drive, the initial
motion is a kink traveling through the pinned skyrmion
lattice. For the square defect array, both models agree
well and show a regime in which the soliton motion is
locked along 45$^\circ$. The atomistic model also
produces locking of a soliton along 30$^\circ$, while
the particle-based model does not. For the triangular
defect array, the atomistic model exhibits soliton
motion locked to 30$^\circ$ over a wide region of
external driving force values. In contrast, the
particle-based model gives soliton motion locked
to 45$^\circ$ over only a small range of external
driving force values. The difference arises because
the nondeforming particle model facilitates meandering
skyrmion orbits while the deformable atomistic model
enables stronger skyrmion-skyrmion interactions that
reduce the meandering. Our results indicate that
soliton motion through pinned skyrmion lattices on a
periodic substrate is a robust effect and could open
the possibility of using solitons as information
carriers. Our results also provide a better understanding
of the regimes for which particle-based models of skyrmions
are best suited.
}

\begin{document}

\maketitle

\section{Introduction}
Solitons are nonlinear wave perturbations \cite{scott_soliton_1973}
that have been observed in a wide range of different science fields,
including
mathematics \cite{manukure_short_2021},
chemistry \cite{heeger_solitons_1988, tolbert_solitons_1992},
magnetism \cite{vizarim_soliton_2022, souza_soliton_2023, galkina_dynamic_2018, slonczewski_dynamics_1972, papanicolaou_dynamics_1991, kosevich_magnon_1978, muhlbauer_skyrmion_2009},
and biology \cite{ciblis_possibility_1997, zhou_biological_1989, davydov_theory_1973}.
Due to their nonlinear spin dynamics,
magnetic systems are particularly well able to stabilize magnetic solitons
or nonlinear magnetic textures,
which can take the form of
magnetic vortices \cite{papanicolaou_dynamics_1991},
magnon drops \cite{kosevich_magnon_1978},
magnetic skyrmions \cite{nagaosa_topological_2013, jiang_direct_2017, gobel_beyond_2021},
hopfions \cite{tai_static_2018, liu_binding_2018, sutcliffe_hopfions_2018, gobel_beyond_2021},
and bimerons \cite{nagase_observation_2021, ohara_reversible_2022, gobel_magnetic_2019, gobel_beyond_2021}.

Solitons or kinks can also be stabilized in assemblies of
particles coupled to a
periodic substrate \cite{Bak82,Reichhardt17}.
When the number of particles equals the number of
potential minima, the system should be free of kinks;
however, if the number of particles
is slightly higher or lower so that the system is off commensuration,
localized kinks appear that depin under applied drive levels which can
be much lower than the drive at which the bulk of the particles depin.
Kink motion at incommensurate fillings on periodic substrates has been
studied for colloidal particles
\cite{bohlein_experimental_2012,Vanossi12a,McDermott13a},
superconducting vortices \cite{Reichhardt97,Gutierrez09} and various
friction models \cite{Braun97,Vanossi13}.
Since skyrmions are also particle-like textures, when they are placed
on a periodic substrate, kinks or solitons could also be stabilized
in the skyrmion lattice.
Recent work by Vizarim {\it et al.}\cite{vizarim_soliton_2022} has shown the
possibility of creating and moving a soliton along
quasi one-dimensional chains
of skyrmions using a particle based model.
%\citet{vizarim_soliton_2022}
After this study, Souza {\it et al.}\cite{souza_soliton_2023}
demonstrated with an atomistic model
that soliton motion along skyrmion chains
is stable and that
the soliton exhibited higher velocities than
free skyrmions.
The work on the quasi-one-dimensional systems opened
the possibility of using soliton or kink motion in
magnetic skyrmions as an information transfer method
for new types of soliton-based devices
employing skyrmions. An open question
is whether soliton motion through skyrmion lattices
remains stable
in more realistic fully two-dimensional systems,
whether the soliton behavior can be captured using both particle-based
and atomistic models,
and where the two models agree or disagree.

Magnetic skyrmions are particle-like topologically
protected magnetic textures \cite{nagaosa_topological_2013, je_direct_2020}
that exhibit many similarities to overdamped particles:
they minimize their repulsive interactions by forming a
triangular array, can be set in motion
by the application of external drives, and can interact
with material defects in a variety of ways
\cite{olson_reichhardt_comparing_2014,Reichhardt17,reichhardt_statics_2022}.
The key difference between skyrmions and
other overdamped particles is the presence of a
non-dissipative Magnus force that causes
skyrmions to move in the absence of disorder at
an angle known as the intrinsic skyrmion Hall angle,
$\theta^\mathrm{int}_\mathrm{sk}$, with respect to the external
driving force
\cite{nagaosa_topological_2013, litzius_skyrmion_2017, iwasaki_universal_2013, jiang_direct_2017, lin_driven_2013, lin_particle_2013}.
In order to simulate all of the degrees of freedom of a skyrmion,
it is necessary to
use computationally expensive models, such as 
the atomistic model \cite{evans_atomistic_2018}, that can capture
behavior such as skyrmion annihilation,
creation and deformation.
To mitigate the computational expense of skyrmion simulations,
Lin {\it et al.}\cite{lin_particle_2013} proposed
a particle-based model for skyrmions that
assumes the skyrmions remain rigid, an approximation that
is valid for low skyrmion densities and low external currents.

Using atomistic simulations and particle based simulations,
we study the dynamical behavior of soliton
motion in magnetic skyrmion lattices on square
and triangular substrates
%compared the results form
%atomistic and particle based models.
just away from commensuration that are subjected to an
external driving force.
%ratio,
%$N_\mathrm{sk}=N_m+1$.
For the square substrate, both models produce
soliton motion along a 45$^\circ$ angle;
however, the atomistic model exhibits an
additional 30$^\circ$
soliton motion that is absent in the particle based model. At higher
drives, the entire skyrmion lattice depins, and the transitions
between the the different soliton and skyrmion flow phases are visible
as changes in the transport curves and average Hall angle of
the kink or skyrmion motion.
For the triangular substrate, we also find regimes of stable soliton
motion, but the models
show substantial differences.
%over a wide range of drives.
The atomistic model exhibits soliton motion
along a 30$^\circ$ angle for a wide range of external
driving forces, whereas the particle model produces
soliton motion along a 45$^\circ$ angle for a small
range of external driving forces.
In the particle model, the trajectory of the soliton is
much more meandering, resulting in flow around an average
angle of 
$45^\circ$, while
in the atomistic model,
the finite size of the skyrmions reduces the amount of
meandering flow that occurs, causing motion along
$30^\circ$ to be more stable.

\section{Methods}

%\begin{figure}[!htb]
\begin{figure}
    \centering
    \includegraphics[width=0.8\columnwidth]{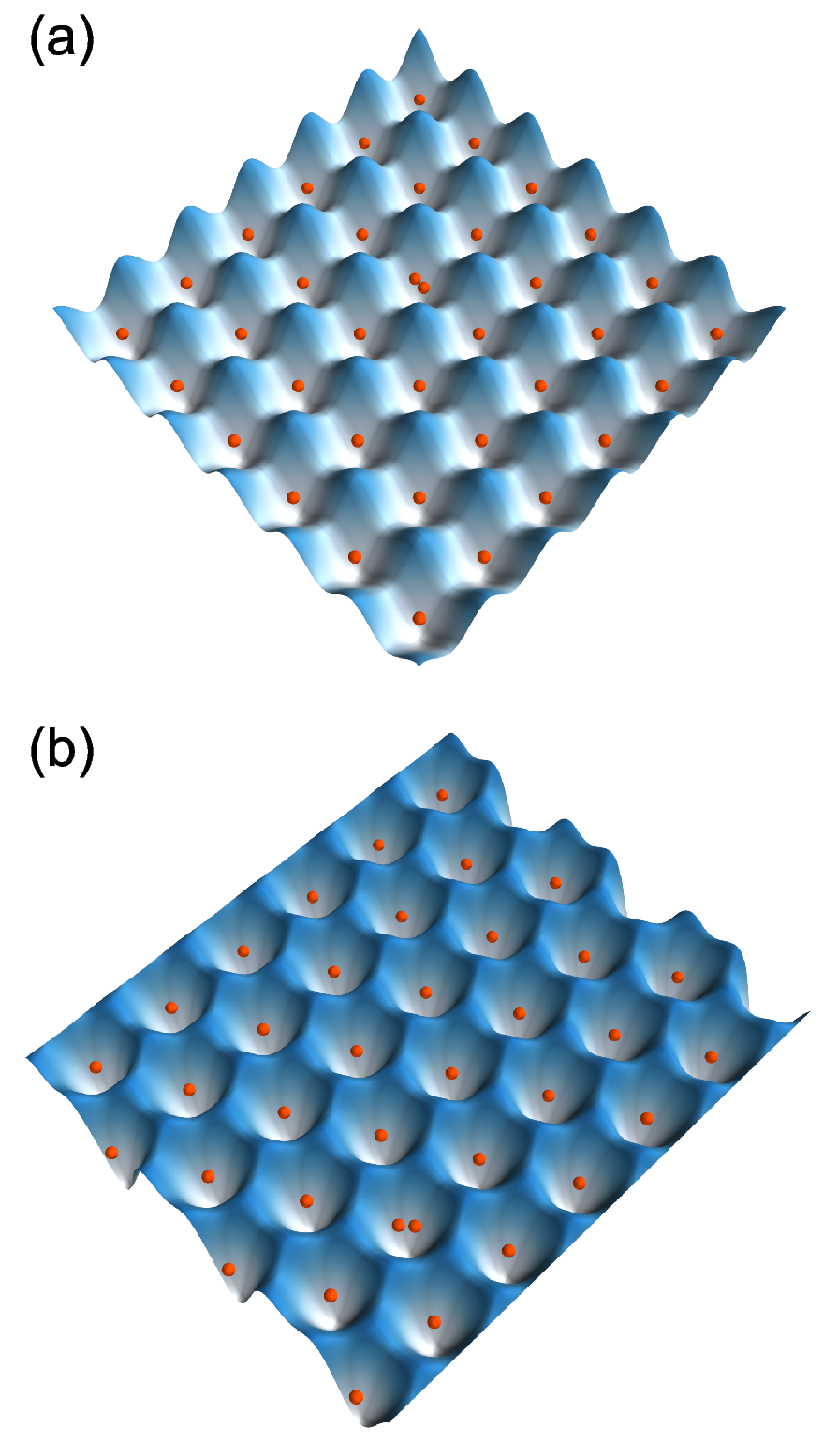}
\caption{
A three-dimensional rendering of the system
for (a) a square substrate potential and (b) a triangular substrate
potential. Each minimum captures one skyrmion, and a single extra
skyrmion has been placed inside the sample in order
to create a kink or incommensuration.
	}
    \label{fig:1}
\end{figure}

We model N\'eel skyrmions in thin films
with a magnetic field applied perpendicular
to the film at zero temperature, $T=0$ K, with
periodic boundary conditions along the $x$ and $y$ directions.
We use two distinct models, the atomistic model
and the particle based model. Common
substrate defect arrangements of $N_m$ defects are used in both
models. The square array of defects is modeled as
$\phi_S(x, y)=\frac{A}{4}\left[\cos\left(\frac{2\pi x}{a_0}\right)+\cos\left(\frac{2\pi y}{a_0}\right) + 2\right]$,
where $A$ is the defect strength and $a_0$ is the substrate
lattice constant. The values of $A$ and $a_0$ are different
between each model and are listed below in the subsections describing
the individual models. 
The triangular array of defects is modeled as
$\phi_T(x, y)=\sum_{i=1}^{3}\frac{A}{2}\left[\cos\left(\frac{2\pi b_i}{a_0}\right) + 1\right]$, with $b_i=x\cos(\theta_i)-y\sin(\theta_i)+a_0/2$ and
$\theta_1=\pi/6$, $\theta_2=\pi/2$, $\theta_3=5\pi/6$.
As in the square array, $A$ is
the defect strength and $a_0$ the substrate lattice constant, with
different
values of $A$ and $a_0$ used for each model.
For both models we choose values of $a_0$ such that
there are $N_m=36$ minima in the defect arrangement.
We select the number of skyrmions $N_\mathrm{sk}$ to be just above
commensuration with the substrate,
$N_\mathrm{sk}=N_m+1=37$.
In fig.~\ref{fig:1}(a,b), we show a three-dimensional rendering
of the square and triangular defect arrays. Each potential minimum captures a
single skyrmion, and we add one additional skyrmion to the sample in
order to create
a kink or soliton
that can depin at a much lower drive
than the commensurate skyrmions. 

\subsection{Atomistic Simulations}
The atomistic model tracks the state of individual
atomic magnetic moments \cite{evans_atomistic_2018}.
The Hamiltonian describing the interactions of
thin films at $T=0$K with an underlying square
atomic arrangement with lattice parameter $a=0.5$ nm is
given by\cite{iwasaki_universal_2013, evans_atomistic_2018, nagaosa_topological_2013}
\begin{eqnarray}\label{eq:1}
  \mathcal{H}=&-\sum_{i, j\in N}J_{ij}\mathbf{m}_i\cdot\mathbf{m}_j
                -\sum_{i, j\in N}\mathbf{D}_{ij}\cdot\left(\mathbf{m}_i\times\mathbf{m}_j\right) \\\nonumber
                &-\sum_i\mu\mathbf{H}\cdot\mathbf{m}_i
                -\sum_{i} K(x_i, y_i)\left(\mathbf{m}_i\cdot\hat{\mathbf{z}}\right)^2 \ .
\end{eqnarray}

The first term on the right side is the exchange interaction
between the nearest neighbors contained in the set $N$,
with an exchange constant of $J_{ij}=J$ between magnetic moments
$i$ and $j$.
The second term is the interfacial Dzyaloshinskii–Moriya
interaction, where $\mathbf{D}_{ij}=D\mathbf{\hat{z}}\times\mathbf{\hat{r}}_{ij}$ is the Dzyaloshinskii–Moriya
vector between magnetic moments $i$ and $j$ and $\mathbf{\hat{r}}_{ij}$
is the
unit distance vector between sites $i$ and $j$.
The third term is the Zeeman interaction with an applied external magnetic
field $\mathbf{H}$.
Here $\mu=\hbar\gamma$ is the magnitude of the magnetic moment
and $\gamma=1.76\times10^{11}$T$^{-1}$s$^{-1}$ is the electron
gyromagnetic ratio. The last term represents the 
perpendicular magnetic anisotropy (PMA) of the sample, where
$x_i$ and $y_i$ are the spatial coordinates of the $i$
magnetic moment. We use
$K(x_i, y_i)=\phi_S(x_i, y_i)$ for a square array of defects
and $K(x_i, y_i)=\phi_T(x_i, y_i)$ for a triangular array of defects.
In ultrathin films,
long-range dipolar interactions can be neglected \cite{paul_role_2020}.

The time evolution of the individual
atomic magnetic moments is given by the LLG
equation \cite{seki_skyrmions_2016, gilbert_phenomenological_2004}

\begin{equation}\label{eq:2}
    \frac{\partial\mathbf{m}_i}{\partial t}=-\gamma\mathbf{m}_i\times\mathbf{H}^\mathrm{eff}_i
                             +\alpha\mathbf{m}_i\times\frac{\partial\mathbf{m}_i}{\partial t}
                             +\frac{pa^3}{2e}\left(\mathbf{j}\cdot\nabla\right)\mathbf{m}_i \ .
\end{equation}
Here
$\mathbf{H}^\mathrm{eff}_i=-\frac{1}{\hbar\gamma}\frac{\partial \mathcal{H}}{\partial \mathbf{m}_i}$
is the effective magnetic field including all interactions from
the Hamiltonian, $\alpha$ is the Gilbert damping, and the last
term is the spin-transfer-torque (STT), where 
$p$ is the spin polarization,
$e$ is the electron charge,
and $\mathbf{j}=j\hat{\mathbf{x}}$ is the applied current density.
The STT current assumes that the conduction electron
spins are always parallel to
the magnetic moments $\mathbf{m}$ \cite{iwasaki_universal_2013,zang_dynamics_2011}, and
the driving force from the STT current \cite{feilhauer_controlled_2020}
acts perpendicular to $\mathbf{j}$.
We fix
$\alpha=0.3$, $J=1$ meV, $D=0.5J$, and $\mu\mathbf{H}=0.6(D^2/J)(-\hat{\mathbf{z}})$.
The resulting skyrmions 
move at an intrinsic skyrmion Hall angle of
$\theta_\mathrm{sk}^\mathrm{int}=64^\circ$ with respect to the driving
force exerted by external currents.
For both
the square and triangular defect arrays
we use $A=0.1J$ and $a_0=14$ nm.
The sample dimensions are 84 nm $\times$ 84 nm for the square
array of defects, and $(2/\sqrt{3})84$ nm $\times$ 84 nm for the
triangular array of defects. The difference in
sample size is required to properly apply boundary conditions.

\subsection{Particle Based Simulations}
The particle based simulations are governed by the
equation of motion\cite{lin_particle_2013}
\begin{equation}\label{eq:3}
    \alpha_d\mathbf{v}_i+\alpha_m\hat{\mathbf{z}}\times\mathbf{v}_i=\sum_{i\neq j}\mathbf{F}_\mathrm{sk}(\mathbf{r}_{ij})+\mathbf{F}_\mathrm{P}(\mathbf{r}_i)+\hat{\mathbf{z}}\times\mathbf{F}_D \ ,
\end{equation}
where $\mathbf{v}_i$ is the velocity of skyrmion $i$.
The first term on the left side is the damping term,
$\alpha_d$, which can be written\cite{iwasaki_universal_2013, everschor-sitte_perspective_2018}
as $\alpha_d=-\alpha\mathcal{D}$,
where $\alpha$ is the Gilbert damping and $\mathcal{D}$ is the
dissipative tensor. The second term is the Magnus force
where the Magnus strength $\alpha_m$ can be
written\cite{iwasaki_universal_2013, everschor-sitte_perspective_2018}
as $\alpha_m=-4\pi Q$, where $Q$ is the skyrmion charge. The
ratio $\alpha_m/\alpha_d$ determines the intrinsic skyrmion
Hall angle $\theta_\mathrm{sk}^\mathrm{int}=\arctan(\alpha_m/\alpha_d)$.
In order to match
the particle based $\theta_\mathrm{sk}$ to the atomistic
$\theta_\mathrm{sk}$,
we use values of $\alpha_m$ and $\alpha_d$ such
that $\theta_\mathrm{sk}^\mathrm{int}=\arctan(\alpha_m/\alpha_d)=64^\circ$.
The first term on the right side of eq.~\ref{eq:3} is the skyrmion-skyrmion
interaction given by
$\mathbf{F}_\mathrm{sk}(\mathbf{r}_{ij})=-U_\mathrm{sk}K_1(r_{ij})\hat{\mathbf{r}}_{ij}$, where $U_\mathrm{sk}=1$ is the
interaction strength and $K_1$ is the modified first order Bessel function.
The second term is the interaction with the underlying
substrate potential, given by
$\mathbf{F}_\mathrm{P}(\mathbf{r}_i)=-\nabla\phi_S(\mathbf{r}_i)$
for the square array of defects and
$\mathbf{F}_\mathrm{P}(\mathbf{r}_i)=-\nabla\phi_T(\mathbf{r}_i)$
for the triangular array of defects.
In both defect arrays, the potential strength is $A=4$ and
the substrate lattice constant is $a_0=6$.
The last term is the interaction with an external drive,
$\mathbf{F}_D=F_D\hat{\mathbf{x}}$, which is in accordance
with the action of an STT current on
magnetic skyrmions \cite{iwasaki_universal_2013, everschor-sitte_perspective_2018, feilhauer_controlled_2020}.
Our simulation box is of size
$36 \times 36$ for the square defect array
and $(2/\sqrt{3})36\times36$ for the triangular defect array.

\section{Square array}

%\begin{figure}[!htb]
\begin{figure}
    \centering
    \includegraphics[width=\columnwidth]{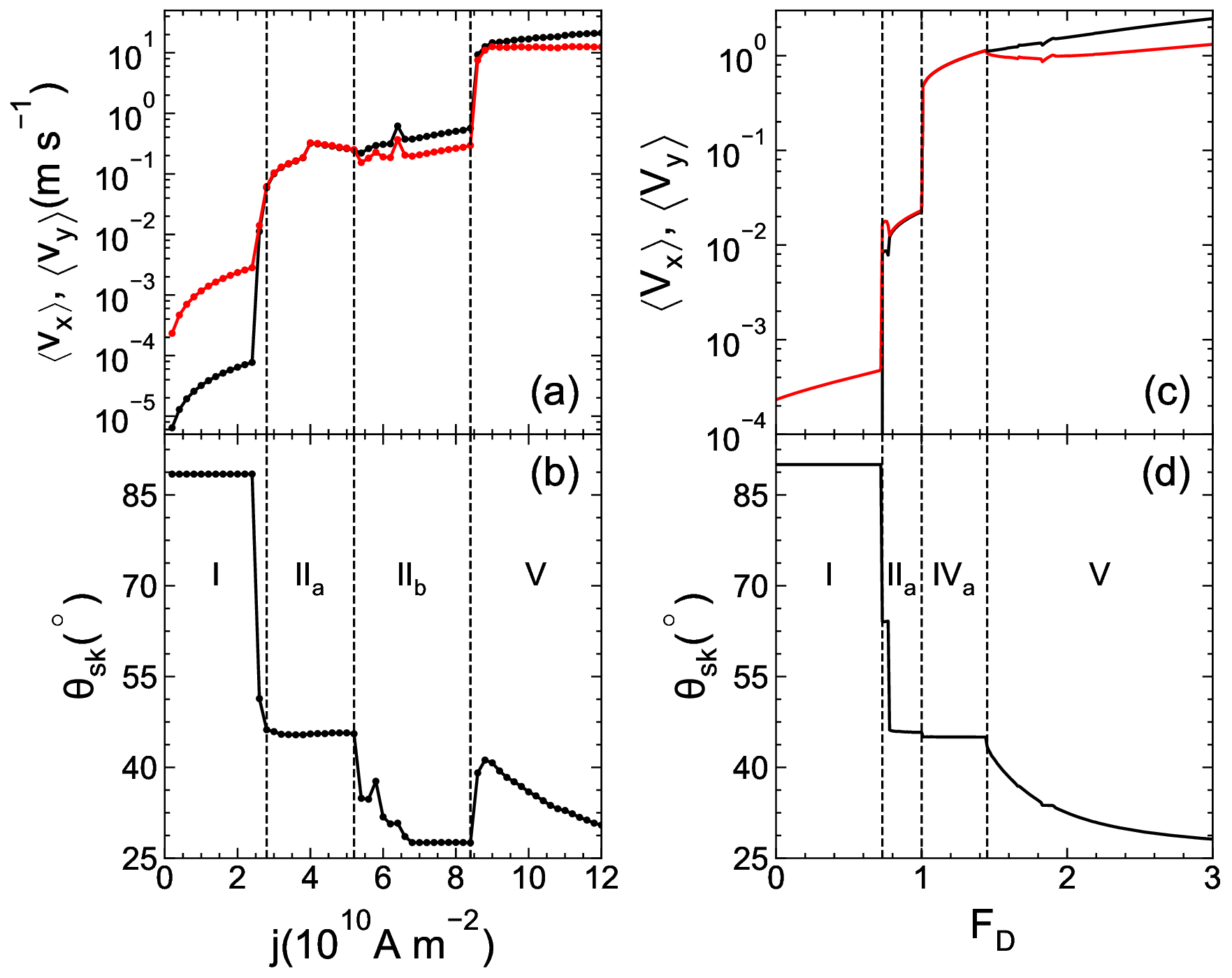}
    \caption{Velocity curves and the corresponding
    skyrmion Hall angle $\theta_\mathrm{sk}$ obtained
    from (a, b) atomistic simulations and (c, d) particle
    based simulations for the square defect array.
    (a) $\left\langle v_x\right\rangle$
    (black) and $\left\langle v_y\right\rangle$ (red)
    vs applied current $j$. (b) The corresponding
    $\theta_\mathrm{sk}$ vs $j$.
    (c) $\left\langle V_x\right\rangle$
    (black) and $\left\langle V_y\right\rangle$ (red)
    vs driving force $F_D$. (d) The corresponding
    $\theta_\mathrm{sk}$ vs $F_D$.
    Dashed lines delimit the different phases.
    I: pinned state.
    II$_a$: soliton motion along $45^\circ$.
    II$_b$: soliton motion along $30^\circ$.
    IV$_a$: all skyrmions moving along $45^\circ$.
    V: all skyrmions moving with no locking angle.
    }
    \label{fig:2}
\end{figure}

\begin{figure}
\centering
\includegraphics[width=\columnwidth]{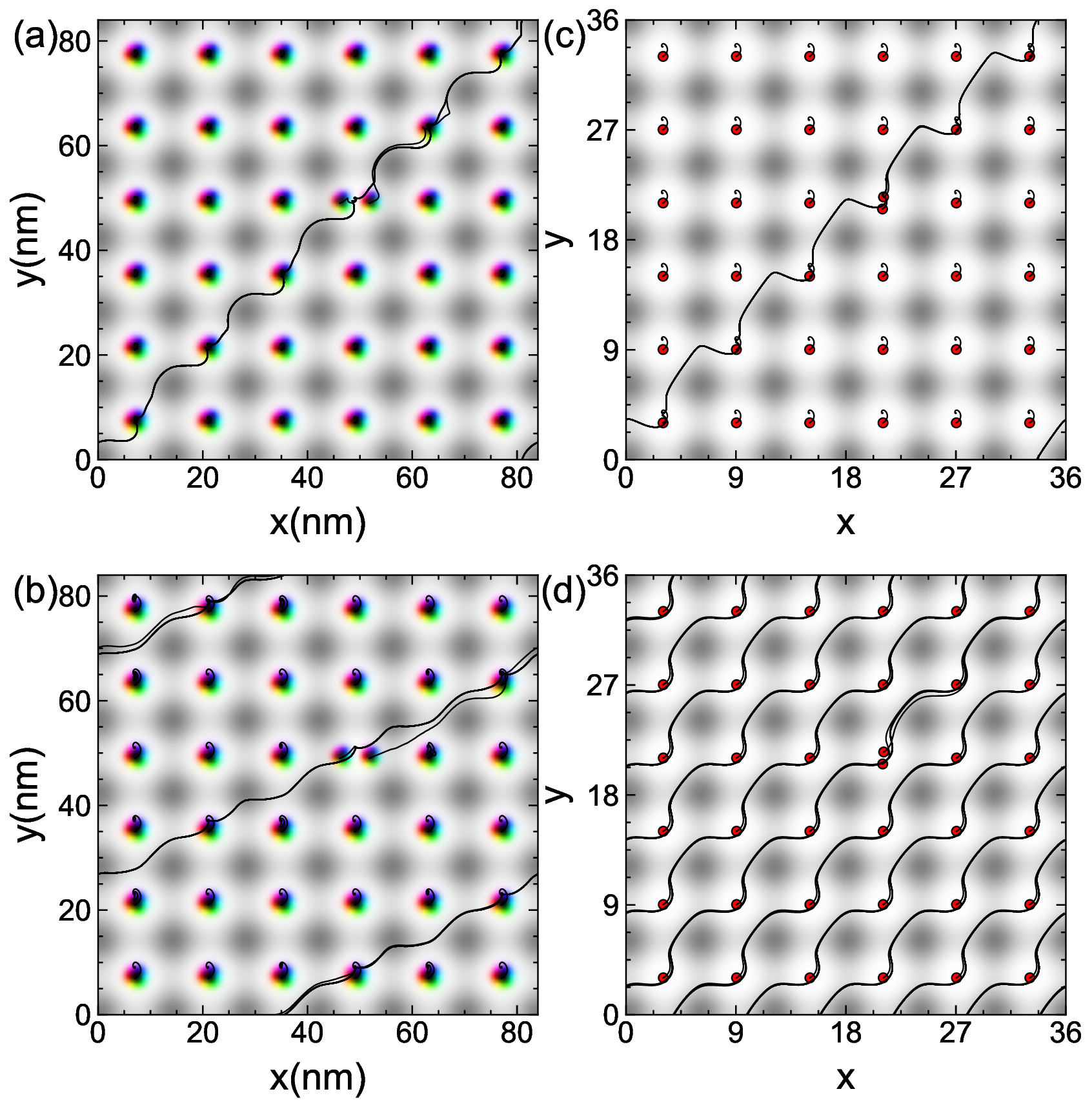}
\caption{Skyrmion trajectories (black lines) and the location of
  substrate minima (white) and maxima (gray) in the square defect array
  for (a, b) atomistic simulations where the skyrmions appear as colored
  disks and (c, d) particle based simulations where the skyrmions are
  drawn as red disks.
  (a) Atomistic simulation at
  $j=4\times10^{10}$A\,m$^{-2}$, corresponding
  to phase II$_a$ from fig.~\ref{fig:2}(a, b).
  (b) Atomistic simulation at
$j=8\times10^{10}$A\,m$^{-2}$, corresponding
  to phase II$_b$ from fig.~\ref{fig:2}(a, b).
  (c) Particle based simulation at
 $F_D=0.8$, corresponding to phase II$_a$ in
  fig.~\ref{fig:2}(c, d).
  (d) Particle based simulation at 
  $F_D=1.2$, corresponding to phase IV$_a$ in
  fig.~\ref{fig:2}(c, d).
  Animations showing the motion of the skyrmions are
  available in the Supplemental Material \cite{Suppl}.
    }
    \label{fig:3}
\end{figure}

We first compare the dynamics of skyrmions
interacting with a square array of defects in atomistic and
particle-based simulations.
In fig.~\ref{fig:2}(a), we plot $\langle v_x\rangle$ and $\langle v_y\rangle$
versus applied current $j$, and in fig.~\ref{fig:2}(b), we show the
corresponding effective $\theta_\mathrm{sk}=\arctan(\langle v_y\rangle/\langle v_x\rangle)$
versus $j$ from the atomistic simulation.
We observe four dynamical phases: a pinned phase I with no
skyrmion motion,
phase II$_a$ where a soliton moves along 45$^\circ$,
phase II$_b$ in which a soliton
moves at 30$^\circ$, and phase V in which
all of the skyrmions depin and move without locking to any direction.
The skyrmion motion in phases
II$_a$ and II$_b$ is illustrated in fig.~\ref{fig:3}(a, b).

%\begin{figure}[!htb]

Figure~\ref{fig:2}(c, d) shows $\langle v_x\rangle$, $\langle v_y\rangle$,
and $\theta_\mathrm{sk}$ versus $F_D$ from particle based simulations
of skyrmions
interacting with a square array of defects.
We again observe four dynamical phases, which are a
pinned phase I,
phase II$_a$ with soliton motion along 45$^\circ$,
phase IV$_a$ in which all of the skyrmions
depin and move along
$\theta_\mathrm{sk}=45^\circ$,
and phase V where all
of the skyrmions have depinned but move without locking to
any direction.
The skyrmion motion in phases II$_a$ and IV$_a$ is illustrated
in fig.~\ref{fig:3}(c, d).

Figure~\ref{fig:3}(a) shows the skyrmion trajectories
during the phase II$_a$ soliton motion along
45$^\circ$ from
the atomistic model.
Here, the extra skyrmion shares an anisotropy minimum with another
skyrmion, and the skyrmion-skyrmion interaction force between the
two lowers the depinning force at which one of the skyrmions can
escape from the minimum and become an interstitial skyrmion.
The interstitial skyrmion moves across the anisotropy landscape
until it reaches another anisotropy
minimum containing a pinned skyrmion.
Through skyrmion-skyrmion interactions,
the interstitial skyrmion pushes the pinned skyrmion
out of the anisotropy minimum and takes up residence in the minimum.
The displaced skyrmion becomes the new
interstitial skyrmion, and this pattern of motion repeats indefinitely.
In fig.~\ref{fig:3}(b) we illustrate the
skyrmion trajectories during the phase II$_b$ soliton
motion along 30$^\circ$ using the atomistic model.
The mechanism
of motion is identical to that observed in fig.~\ref{fig:3}(a),
but now the interstitial skyrmion encounters a pinned
skyrmion along the 30$^\circ$ angle line instead of
the 45$^\circ$ angle line.

Figure~\ref{fig:3}(c) shows the skyrmion trajectories during the
phase II$_a$ soliton motion at 45$^\circ$ obtained from
the particle based model. The
behavior is identical to that found in the atomistic model,
where the skyrmion-skyrmion interactions cause the extra skyrmion
trapped inside a potential minimum to depin at low
$F_D$.
Since the skyrmion is treated as
a point particle, the trajectory differs in detail from the
atomistic trajectory shown
in fig.~\ref{fig:3}(a); however,
the mechanism of exchange between interstitial and pinned skyrmions
remains the same.
In fig.~\ref{fig:3}(d) we show
the skyrmion trajectories in phase IV$_a$ from the
particle based model
where all of the skyrmions have depinned and
move along 45$^\circ$.
Here the extra skyrmion does not play
a major role, since all
of the skyrmions are 
interacting with the potential in an ordered manner and following
a 45$^\circ$ trajectory. The skyrmion lattice is similar
to a moving crystal but contains a localized defect produced by
the extra skyrmion.

We note that previous work on skyrmions
moving over a two-dimensional square periodic  
substrate under an increasing drive
showed a directional locking effect in which
the skyrmion motion locked to particular symmetry angles of the 
substrate \cite{Reichhardt15a,Reichhardt22b}.
Continuum models for individual
skyrmions on antidot lattices
also produce similar directional 
locking
\cite{feilhauer_controlled_2020,Sun23}. 
The results we describe here are different
in that the motion is not of individual
continuously moving skyrmions
but of kink traveling through a skyrmion lattice,
so the locking is a collective effect instead of a single
particle effect.
Additionally, the applied drive needed
to produce kink motion is substantially lower
that the drive at which an isolated skyrmion would depin from
the substrate potential minimum. The reduction in the
driving
threshold would be particularly useful for applications. 

\section{Triangular array}

%\begin{figure}[!htb]
\begin{figure}
    \centering
    \includegraphics[width=\columnwidth]{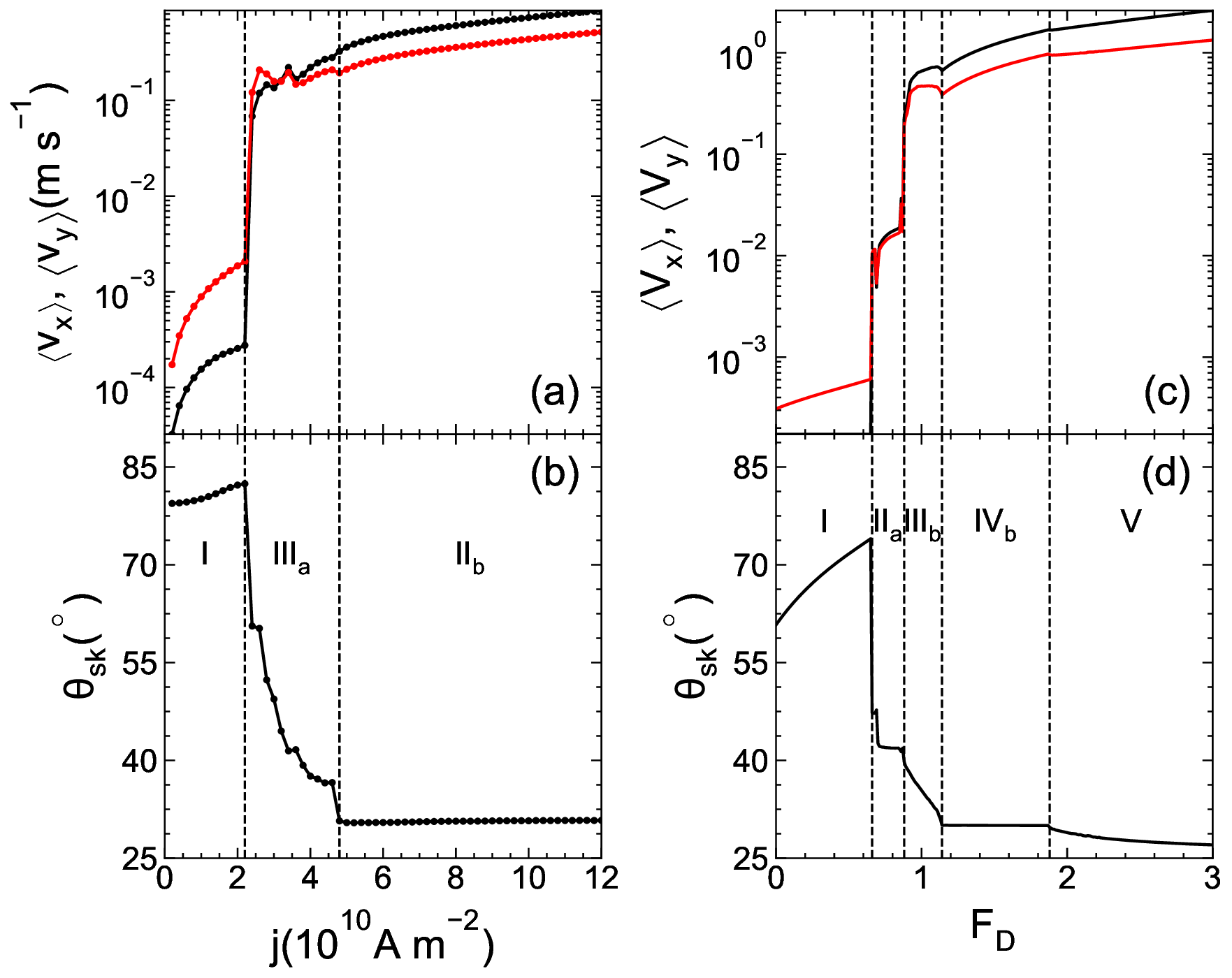}
    \caption{Velocity curves and the corresponding
    skyrmion Hall angle $\theta_\mathrm{sk}$ obtained from
    (a, b) atomistic simulations and (c, d) particle
    based simulations for the triangular
    defect array.
    (a) $\left\langle v_x\right\rangle$ (black)
    and $\left\langle v_y\right\rangle$ (red) vs
    applied current $j$.
    (b) The corresponding $\theta_\mathrm{sk}$ vs $j$.
    (c) $\left\langle V_x\right\rangle$ (black)
    and $\left\langle V_y\right\rangle$ (red) vs
    driving force $F_D$.
    (d) The corresponding $\theta_\mathrm{sk}$ vs
    $F_D$.
    Dashed lines delimit different phases:
    I: pinned state.
    II$_a$: soliton motion along 45$^\circ$.
    II$_b$: soliton motion along 30$^\circ$.
    III$_a$: disordered soliton motion.
    III$_b$: completely disordered flow.
    IV$_b$: all skyrmions moving along 30$^\circ$.
    V: all skyrmions moving without locking to any direction.}
    %CIJOL WORKING
    %I pinned II disordered soliton III 30^o soliton.
    %I pinned II 45^o soliton III transition IV all moving at 30^c V all moving
    \label{fig:4}
\end{figure}

\begin{figure}
    \centering
    \includegraphics[width=\columnwidth]{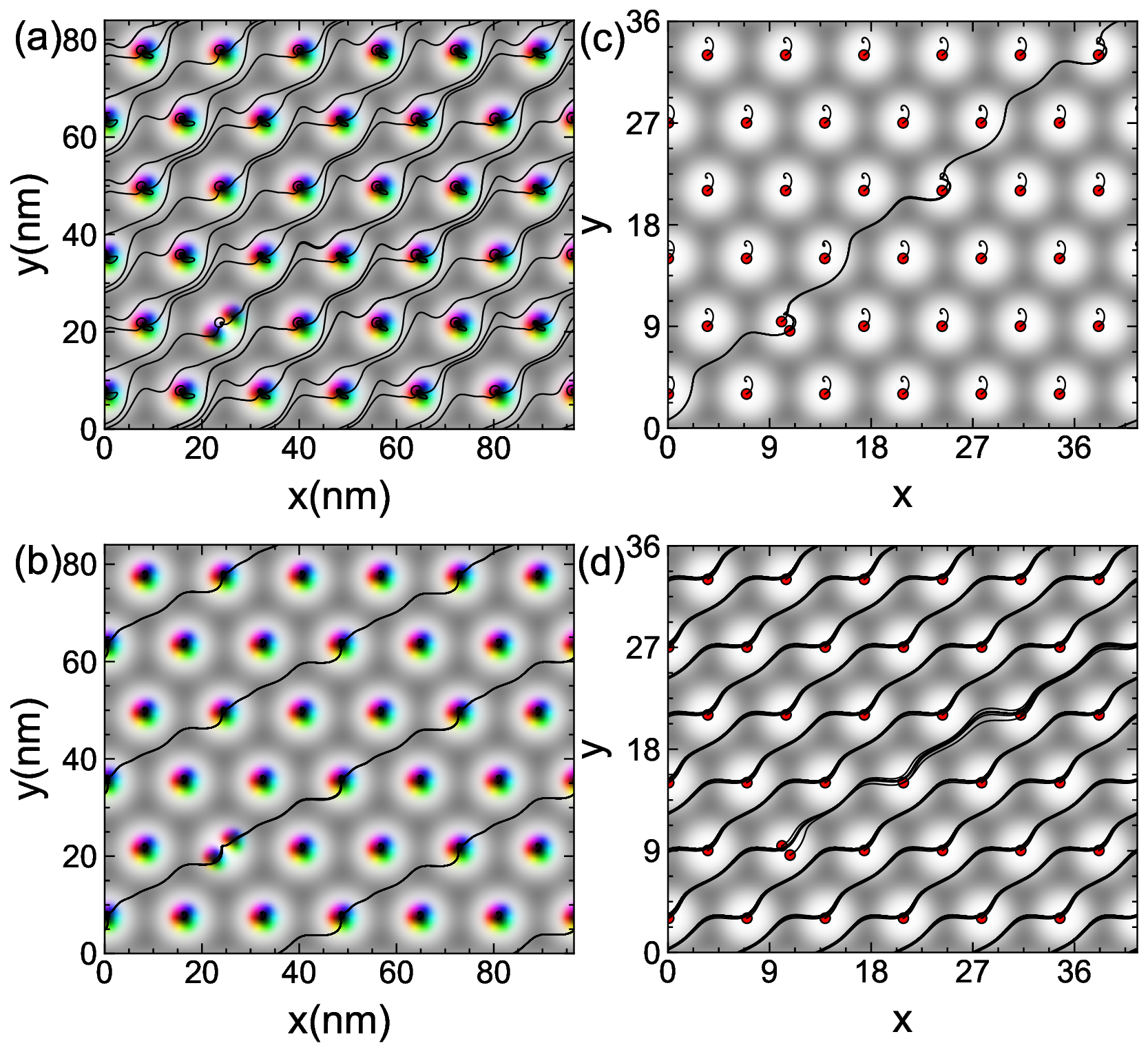}
    \caption{Skyrmion trajectories (black lines) and the location
    of substrate minima (white) and maxima (gray) in the
    triangular defect array for
    (a, b) atomistic simulations where the skyrmions appear as colored disks
    and
    (c, d) particle based simulations where the skyrmions are drawn as
    red disks.
    (a) Atomistic simulation at
    $j=4\times10^{10}$A\,m$^{-2}$, corresponding to 
    phase III$_a$ in fig.~\ref{fig:4}(a, b).
    (b) Atomistic simulation at
    $j=8\times10^{10}$A\,m$^{-2}$, corresponding to
    phase II$_b$ in fig.~\ref{fig:4}(a, b).
    (c) Particle based simulation at
    $F_D=0.8$, corresponding to phase II$_a$ in
    fig.~\ref{fig:4}(c, d).
    (d) Particle based simulation at $F_D=1.5$,
    corresponding to phase IV$_b$
    in fig.~\ref{fig:4}(c, d).
    Animations showing the skyrmion motion
    are available in the Supplemental Material \cite{Suppl}.}
    \label{fig:5}
\end{figure}

We next compare the dynamical behavior of skyrmions on
a triangular defect array
in atomistic simulations and particle based simulations.
For the atomistic simulations,
fig.~\ref{fig:4}(a) shows $\langle v_x\rangle$ and $\langle v_y\rangle$
versus $j$, while in fig.~\ref{fig:4}(b) we plot the corresponding
$\theta_\mathrm{sk}$ versus $j$.
We observe three dynamical phases.
At low drives, we find a pinned phase I.
There is a transitional phase III$_a$ of disordered soliton
motion, where soliton transport occurs but does not follow a
well defined direction.
At higher drives, phase II$_b$ appears in which the soliton
moves along
30$^\circ$. Over the range of $j$ values that we consider, we
never observe a completely depinned state, but we expect that for
larger values of $j$,
the pinned skyrmions would eventually escape from
the anisotropy minima and move throughout the sample.
The skyrmion trajectories
in phases III$_a$ and II$_b$ are illustrated
in fig.~\ref{fig:5}(a, b).

%\begin{figure}[!htb]

In fig.~\ref{fig:4}(c), we plot $\langle V_x\rangle$ and
$\langle V_y\rangle$ versus $F_D$ for a particle based simulation
of skyrmions driven over a triangular defect array, and
in fig.~\ref{fig:4}(d) we show the corresponding
$\theta_\mathrm{sk}$ versus $F_D$.
Five dynamical phases appear.
At low drives, we find a pinned phase I.
Phase II$_a$ is soliton motion along
45$^\circ$, and it is followed by a transitional phase III$_b$ in which
a small number of skyrmions are present simultaneously and move
chaotically across the system.
In phase IV$_b$,
all of the skyrmions are moving along 30$^\circ$,
and phase V consists of
all of the skyrmions moving without locking to any direction.
Illustrations of the skyrmion trajectories
for phases II$_a$ and IV$_b$ appear in fig.~\ref{fig:5}(c, d).

Unlike what we found for the square defect array,
here the atomistic model and the particle based model
do not exhibit good agreement.
A pinned phase is present in both cases, but as we increase
the drive, in the atomistic simulations
we observe a wide phase II$_b$ 30$^\circ$ soliton
motion over the
range
$4.8\times10^{10}\mathrm{A\,m}^{-2}\leq j\leq 12\times10^{10}\mathrm{A\,m}^{-2}$.
In comparison, for the particle based simulations
there is a small window of phase II$_a$
45$^\circ$ soliton motion over the range
$0.66\leq F_D\leq 0.88$.
The phase II$_b$ 30$^\circ$ angle soliton motion is
absent in the particle based simulations.
This large difference in behavior is likely due to the fact that
in the atomistic simulations, the skyrmion has a
finite size and is able to deform, as is visible by
comparing the phase II$_a$ flow in fig.~\ref{fig:3}(a)
to that in fig.~\ref{fig:3}(c), or
comparing the phase II$_b$ flow in fig.~\ref{fig:5}(b)
to the phase II$_a$ flow in fig.~\ref{fig:5}(c).
For higher values of $F_D$, the behavior of the atomistic and
particle based simulations diverge significantly.
The particle based simulation
produces phase IV$_b$ flow in which
all of the skyrmions move along 30$^\circ$, followed by
phase V flow in which all of the skyrmions move without locking
to a particular direction,
while the atomistic simulations remain trapped in phase II$_b$
with soliton motion along 30$^\circ$.

Figure~\ref{fig:5}(a) shows the trajectories of skyrmions
interacting
with a triangular defect array from atomistic
simulations performed at $j=4\times10^{10}$A\,m$^{-2}$, corresponding to
phase III$_a$ in fig.~\ref{fig:4}(a, b). The soliton mechanism of motion
described previously still occurs here, but the soliton does not
lock to any angle and gradually works its way all around the sample.
When we increase the external current to $j=8\times10^{10}$A\,m$^{-2}$,
corresponding to phase II$_b$ in fig.~\ref{fig:4}(a, b),
the skyrmions move as illustrated in fig.~\ref{fig:5}(b).
Again, we observe a soliton motion, but unlike what was shown
in fig.~\ref{fig:5}(a),
the motion occurs along a well defined angle of 30$^\circ$.
This motion
appears
over the range $4.8\times10^{10}\mathrm{A\,m}^{-2}\leq j\leq 12\times10^{10}\mathrm{A\,m}^{-2}$.

In fig.~\ref{fig:5}(c) we illustrate the skyrmion trajectories for motion
over a triangular defect array from
particle based simulations
with $F_D=0.8$, corresponding to the phase II$_a$ flow
in fig.~\ref{fig:4}(c, d).
For this value of
$F_D$, a soliton moves across the sample at 
45$^\circ$.
We observe the 45$^\circ$ soliton motion only
in the particle based simulations and do not find it in the
atomic simulations.
When we increase $F_D$ to $F_D=1.5$,
corresponding to phase IV$_b$ in fig.~\ref{fig:4}(c, d),
all of the skyrmions flow along 30$^\circ$,
as shown in fig.~\ref{fig:5}(d).
Similar to what we found in
fig.~\ref{fig:3}(d), the skyrmion lattice travels as a moving
crystal that contains
a localized defect produced by
the extra skyrmion.

The difference in the angle of soliton motion on a triangular
defect array between the atomistic and particle based models
can be explained
by the finite skyrmion size in
the atomistic simulations.
The barrier between anisotropy minima is enhanced when the skyrmion
has a finite size, whereas in the particle based simulations,
the pointlike nature of the skyrmions gives a reduced barrier
between anisotropy minima.
The larger barrier potential created by the finite
skyrmion size can be observed
by comparing the
trajectory of an interstitial skyrmion as it
passes between two anisotropy minima;
the trajectories in the atomistic simulation
exhibit fewer meanders compared to the trajectories in the particle
based simulations.
The increase in the barrier potential
forces the atomistic interstitial
skyrmion to move following the 30$^\circ$ angle imposed by the
triangular lattice,
whereas the interstitial skyrmion in the
particle based model can travel along a wider range of paths
between substrate minima.
To compensate for the overly unconstrained mobility of
skyrmions the particle
based model, the barrier between potential minima can be increased
either by
increasing
the skyrmion-skyrmion interaction strength $U_\mathrm{sk}$ or
reducing the lattice constant of the substrate potential.
In principle, it may be possible to identify simulation parameters
for the particle model of the triangular substrate that would match
the behavior of the atomistic simulations and compensate for the
rigidity and vanishing size of the particle-based skyrmion model.

\section{Summary}

We compared the results of atomistic simulations and particle based
simulations of soliton motion for skyrmion assemblies just past
commensuration on both square and triangular substrates.
For the square array, both models agree well at low and high
drives, but differ for intermediate drives.
The atomistic model produces
a pinned phase, soliton motion along 45$^\circ$,
soliton motion along 30$^\circ$, and unlocked flow of all skyrmions.
The soliton motion proceeds via the replacement
by an interstitial skyrmion of
a pinned skyrmion in an anisotropy minimum, with the depinned skyrmion
becoming the new interstitial skyrmion.
The particle based model produces
a pinned phase, soliton motion along 45$^\circ$,
a phase where all of the skyrmions move
along 45$^\circ$, and unlocked flow of all skyrmions.
The particle based model
does not exhibit the 30$^\circ$ soliton motion
found in the atomistic
simulations, and the trajectories in the particle based model meander
more than the atomistic model trajectories due to the rigidity and
vanishing size of the particle based skyrmions.
For both models, the different dynamic phases are visible
as signatures
in the velocity-force curves, skyrmion Hall angle, and skyrmion trajectories. 

The atomistic and particle based models do not agree well on the
motion of skyrmions over
a triangular defect array.
The atomistic model produces
a pinned phase, a transitional phase in which a soliton moves with
no well defined angle,
and a regime of soliton motion at 
30$^\circ$ that
spans a wide range of external drive values.
The particle based model exhibits 
a pinned phase, soliton motion along 45$^\circ$,
a transitional phase in which all skyrmions participate in disordered
soliton motion,
a phase in which all of the skyrmions move
along 30$^\circ$, and unlocked motion of all of the skyrmions.
Here only the pinned phase is common
between the two models.

Our results provide a better understanding of the
regimes in which the particle model is a good or a poor approximation
for the skyrmion motion.
We argue that it can be possible to mitigate the approximations
made in the particle based model by adjusting the strength of the
interactions between the skyrmions or modifying the lattice constant
of the substrate.
Our results will be beneficial for determining how to
control skyrmion soliton 
motion using a combination of anisotropy trapping and
external driving.

\acknowledgments
We gratefully acknowledge the support of the U.S. Department of
Energy through the LANL/LDRD program for this work.
This work was supported by the US Department of Energy through the Los Alamos National Laboratory. Los
Alamos National Laboratory is operated by Triad National Security, LLC, for the National Nuclear Security
Administration of the U. S. Department of Energy (Contract No. 892333218NCA000001). 
J.C.B.S acknowledges funding from Fundação de Amparo à Pesquisa do Estado de São Paulo - FAPESP (Grant 2023/17545-1).
We would like to thank Dr. Felipe F. Fanchini for providing the computational resources used in this work. 
These resources were funded by the Fundação de Amparo à Pesquisa do Estado de São Paulo - FAPESP (Grant: 2021/04655-8).

\bibliographystyle{eplbib}
\bibliography{mybib}

\begin{thebibliography}{10}
\expandafter\ifx\csname url\endcsname\relax\def\url#1{\texttt{#1}}\fi

\bibitem{scott_soliton_1973}
\Name{Scott A.~C., Chu F. Y.~F. \and McLaughlin D.~W.} \REVIEW{Proceedings of
  the IEEE}{61}{1973}{1443}.

\bibitem{manukure_short_2021}
\Name{Manukure S. \and Booker T.} \REVIEW{Partial Diff. Eq. Appl.
  Math.}{4}{2021}{100140}.

\bibitem{heeger_solitons_1988}
\Name{Heeger A.~J., Kivelson S., Schrieffer J.~R. \and Su W.~P.} \REVIEW{Rev.
  Mod. Phys.}{60}{1988}{781}.

\bibitem{tolbert_solitons_1992}
\Name{Tolbert L.~M.} \REVIEW{Accounts Chem. Res.}{25}{1992}{561}.

\bibitem{vizarim_soliton_2022}
\Name{Vizarim N.~P., Souza J. C.~B., Reichhardt C. J.~O., Reichhardt C.,
  Milo{\v s}evi{\' c} M.~V. \and Venegas P.~A.} \REVIEW{Phys. Rev.
  B}{105}{2022}{224409}.

\bibitem{souza_soliton_2023}
\Name{Souza J. C.~B., Vizarim N.~P., Reichhardt C. J.~O., Reichhardt C. \and
  Venegas P.~A.} \REVIEW{J. Mag. Mag. Mater.}{587}{2023}{171280}.

\bibitem{galkina_dynamic_2018}
\Name{Galkina E.~G. \and Ivanov B.~A.} \REVIEW{Low Temp. Phys.}{44}{2018}{618}.

\bibitem{slonczewski_dynamics_1972}
\Name{Slonczewski J.~C.} \REVIEW{AIP Conf. Proc.}{5}{1972}{170}.

\bibitem{papanicolaou_dynamics_1991}
\Name{Papanicolaou N. \and Tomaras T.~N.} \REVIEW{Nucl. Phys.
  B}{360}{1991}{425}.

\bibitem{kosevich_magnon_1978}
\Name{Kosevich A.~M., Ivanov B.~A. \and Kovalev A.~S.} \REVIEW{J. Phys.
  Colloq.}{39}{1978}{C6}.

\bibitem{muhlbauer_skyrmion_2009}
\Name{M{\" u}hlbauer S., Binz B., Jonietz F., Pfleiderer C., Rosch A., Neubauer
  A., Georgii R. \and B{\" o}ni P.} \REVIEW{Science}{323}{2009}{915}.

\bibitem{ciblis_possibility_1997}
\Name{Ciblis P. \and Cosic I.} \REVIEW{J. Theor. Biol.}{184}{1997}{331}.

\bibitem{zhou_biological_1989}
\Name{Zhou G.-P.} \REVIEW{Phys. Scripta}{40}{1989}{698}.

\bibitem{davydov_theory_1973}
\Name{Davydov A.~S.} \REVIEW{J. Theor. Biol.}{38}{1973}{559}.

\bibitem{nagaosa_topological_2013}
\Name{Nagaosa N. \and Tokura Y.} \REVIEW{Nature Nanotechnol.}{8}{2013}{899}.

\bibitem{jiang_direct_2017}
\Name{Jiang W., Zhang X., Yu G., Zhang W., Wang X., Jungfleisch M.~B., Pearson
  J.~E., Cheng X., Heinonen O., Wang K.~L., Zhou Y., Hoffmann A. \and
  te~Velthuis S. G.~E.} \REVIEW{Nature Phys.}{13}{2017}{162}.

\bibitem{gobel_beyond_2021}
\Name{G{\" o}bel B., Mertig I. \and Tretiakov O.~A.} \REVIEW{Phys.
  Rep.}{895}{2021}{1}.

\bibitem{tai_static_2018}
\Name{Tai J.-S.~B. \and Smalyukh I.~I.} \REVIEW{Phys. Rev.
  Lett.}{121}{2018}{187201}.

\bibitem{liu_binding_2018}
\Name{Liu Y., Lake R.~K. \and Zang J.} \REVIEW{Phys. Rev. B}{98}{2018}{174437}.

\bibitem{sutcliffe_hopfions_2018}
\Name{Sutcliffe P.} \REVIEW{J. Phys. A: Math. Theor.}{51}{2018}{375401}.

\bibitem{nagase_observation_2021}
\Name{Nagase T., So Y.-G., Yasui H., Ishida T., Yoshida H.~K., Tanaka Y.,
  Saitoh K., Ikarashi N., Kawaguchi Y., Kuwahara M. \and Nagao M.}
  \REVIEW{Nature Commun.}{12}{2021}{3490}.

\bibitem{ohara_reversible_2022}
\Name{Ohara K., Zhang X., Chen Y., Kato S., Xia J., Ezawa M., Tretiakov O.~A.,
  Hou Z., Zhou Y., Zhao G., Yang J. \and Liu X.} \REVIEW{Nano
  Lett.}{22}{2022}{8559}.

\bibitem{gobel_magnetic_2019}
\Name{G{\" o}bel B., Mook A., Henk J., Mertig I. \and Tretiakov O.~A.}
  \REVIEW{Phys. Rev. B}{99}{2019}{060407}.

\bibitem{Bak82}
\Name{Bak P.} \REVIEW{Rep. Prog. Phys.}{45}{1982}{587}.

\bibitem{Reichhardt17}
\Name{Reichhardt C. \and Reichhardt C. J.~O.} \REVIEW{Rep. Prog.
  Phys.}{80}{2017}{026501}.

\bibitem{bohlein_experimental_2012}
\Name{Bohlein T. \and Bechinger C.} \REVIEW{Phys. Rev.
  Lett.}{109}{2012}{58301}.

\bibitem{Vanossi12a}
\Name{Vanossi A. \and Tosatti E.} \REVIEW{Nature Mater.}{11}{2012}{97}.

\bibitem{McDermott13a}
\Name{McDermott D., Amelang J., Reichhardt C. J.~O. \and Reichhardt C.}
  \REVIEW{Phys. Rev. E}{88}{2013}{062301}.

\bibitem{Reichhardt97}
\Name{Reichhardt C., Olson C.~J. \and Nori F.} \REVIEW{Phys. Rev.
  Lett.}{78}{1997}{2648}.

\bibitem{Gutierrez09}
\Name{Gutierrez J., Silhanek A.~V., Van~de Vondel J., Gillijns W. \and
  Moshchalkov V.~V.} \REVIEW{Phys. Rev. B}{80}{2009}{140514}.

\bibitem{Braun97}
\Name{Braun O.~M., Dauxois T., Paliy M.~V. \and Peyrard M.} \REVIEW{Phys. Rev.
  Lett.}{78}{1997}{1295}.

\bibitem{Vanossi13}
\Name{Vanossi A., Manini N., Urbakh M., Zapperi S. \and Tosatti E.}
  \REVIEW{Rev. Mod. Phys.}{85}{2013}{529}.

\bibitem{je_direct_2020}
\Name{Je S.-G., Han H.-S., Kim S.~K., Montoya S.~A., Chao W., Hong I.-S.,
  Fullerton E.~E., Lee K.-S., Lee K.-J., Im M.-Y. \and Hong J.-I.} \REVIEW{ACS
  Nano}{14}{2020}{3251}.

\bibitem{olson_reichhardt_comparing_2014}
\Name{Olson~Reichhardt C.~J., Lin S.~Z., Ray D. \and Reichhardt C.}
  \REVIEW{Physica C}{503}{2014}{52}.

\bibitem{reichhardt_statics_2022}
\Name{Reichhardt C., Reichhardt C. J.~O. \and Milo{\v s}evi{\' c} M.}
  \REVIEW{Rev. Mod. Phys.}{94}{2022}{035005}.

\bibitem{litzius_skyrmion_2017}
\Name{Litzius K., Lemesh I., Kr\"uger B., Bassirian P., Caretta L., Richter K.,
  B\"uttner F., Sato K., Tretiakov O.~A., F\"orster J., Reeve R.~M., Weigand
  M., Bykova I., Stoll H., Sch\"utz G., Beach G. S.~D. \and Kl\"aui M.}
  \REVIEW{Nature Phys.}{13}{2017}{170}.

\bibitem{iwasaki_universal_2013}
\Name{Iwasaki J., Mochizuki M. \and Nagaosa N.} \REVIEW{Nature
  Commun.}{4}{2013}{1463}.

\bibitem{lin_driven_2013}
\Name{Lin S.-Z., Reichhardt C., Batista C.~D. \and Saxena A.} \REVIEW{Phys.
  Rev. Lett.}{110}{2013}{207202}.

\bibitem{lin_particle_2013}
\Name{Lin S.-Z., Reichhardt C., Batista C.~D. \and Saxena A.} \REVIEW{Phys.
  Rev. B}{87}{2013}{214419}.

\bibitem{evans_atomistic_2018}
\Name{Evans R. F.~L.} \Book{Atomistic {Spin} {Dynamics}} in \Book{Handbook of
  {Materials} {Modeling}: {Applications}: {Current} and {Emerging}
  {Materials}}, edited by \Name{Andreoni W. \and Yip S.} (Springer
  International Publishing) 2018 pp. 1--23.

\bibitem{paul_role_2020}
\Name{Paul S., Haldar S., von Malottki S. \and Heinze S.} \REVIEW{Nature
  Commun.}{11}{2020}{4756}.

\bibitem{seki_skyrmions_2016}
\Name{Seki S. \and Mochizuki M.} \Book{Skyrmions in {Magnetic} {Materials}}
  (Springer International Publishing) 2016.

\bibitem{gilbert_phenomenological_2004}
\Name{Gilbert T.~L.} \REVIEW{IEEE Trans. Mag.}{40}{2004}{3443}.

\bibitem{zang_dynamics_2011}
\Name{Zang J., Mostovoy M., Han J.~H. \and Nagaosa N.} \REVIEW{Phys. Rev.
  Lett.}{107}{2011}{136804}.

\bibitem{feilhauer_controlled_2020}
\Name{Feilhauer J., Saha S., Tobik J., Zelent M., Heyderman L.~J. \and
  Mruczkiewicz M.} \REVIEW{Phys. Rev. B}{102}{2020}{184425}.

\bibitem{everschor-sitte_perspective_2018}
\Name{Everschor-Sitte K., Masell J., Reeve R.~M. \and Kl{\" a}ui M.} \REVIEW{J.
  Appl. Phys.}{124}{2018}{240901}.

\bibitem{Suppl}
\Name{for~supplementary videos. S. S.~M.}

\bibitem{Reichhardt15a}
\Name{Reichhardt C., Ray D. \and Reichhardt C. J.~O.} \REVIEW{Phys. Rev.
  B}{91}{2015}{104426}.

\bibitem{Reichhardt22b}
\Name{Reichhardt C. \and Reichhardt C. J.~O.} \REVIEW{Phys. Rev.
  B}{105}{2022}{214437}.

\bibitem{Sun23}
\Name{Sun W., Wang W., Yang C., Li X., Li H., Huang S. \and Cheng Z.}
  \REVIEW{Phys. Rev. B}{107}{2023}{184439}.

\end{thebibliography}

\end{document}